\documentclass{article}

\usepackage[preprint, nonatbib]{neurips_2026} 
\usepackage[numbers, compress]{natbib} 

\usepackage[utf8]{inputenc} 
\usepackage[T1]{fontenc}    
\usepackage{hyperref}       
\usepackage{url}            
\usepackage{booktabs}       
\usepackage{amsfonts}       
\usepackage{nicefrac}       
\usepackage{microtype}      
\usepackage{xcolor}         
\usepackage{pifont}

\usepackage{amsmath}
\usepackage{graphicx}
\graphicspath{{figures/}}
\usepackage{subcaption}
\usepackage{multirow}
\usepackage{makecell}
\usepackage{algorithm}
\usepackage{algorithmic}
\usepackage{wrapfig}
\usepackage{enumitem}
\usepackage{listings}

\usepackage{xspace}
\usepackage{tabularx}
\usepackage{bbm}

\newcommand{\sbuf}{\texttt{shared\_buffers}}

\newcommand{\cct}{\texttt{checkpoint\_completion\_target}}
\newcommand{\rpc}{\texttt{random\_page\_cost}}

\title{
A Case for Agentic Tuning: \\ From Documentation to Action in PostgreSQL
}
\author{%
  Hongyu Lin\thanks{These authors contributed equally.}\\
  Institute of Software, Chinese Academy of Sciences\\
  University of Chinese Academy of Sciences\\
  \texttt{hongyu2021@iscas.ac.cn}
  \And
  Mingyu Li\footnotemark[1]\\
  Key Laboratory of System Software (Chinese Academy of Sciences)\\
  Institute of Software, Chinese Academy of Sciences\\
  University of Chinese Academy of Sciences\\
  \texttt{limingyu@ios.ac.cn}
  \And
  Weichen Zhang\\
  Beihang University\\
  \texttt{summer\_05\_05@buaa.edu.cn}
  \And
  Yihang Lou\\
  Peking University\\
  \texttt{yihanglou@pku.edu.cn}
  \AND
  Mingjie Xing\\
  Institute of Software, Chinese Academy of Sciences\\
  University of Chinese Academy of Sciences\\
  \texttt{mingjie@iscas.ac.cn}
  \And
  Yanjun Wu\\
  Institute of Software, Chinese Academy of Sciences\\
  University of Chinese Academy of Sciences\\
  \texttt{yanjun@iscas.ac.cn}
  \And
  Haibo Chen\\
  Key Laboratory of System Software (Chinese Academy of Sciences)\\
  Institute of Software, Chinese Academy of Sciences\\
  Shanghai Jiao Tong University\\
  \texttt{haibochen@ios.ac.cn}
}
\newenvironment{myitemize}%
  {\begin{list}{\labelitemi}{\itemsep3pt \topsep3pt \parsep0.00in
  \partopsep=3pt \leftmargin1.2em}}%
  {\end{list}}

\newcommand{\name}{\textsc{PerfEvolve}\xspace}

\newcommand{\heading}[1]{\vspace{0.5em}\noindent\textbf{#1}}

\begin{document}

\maketitle

\begin{abstract}

Documentation has long guided computer system tuning by distilling expert knowledge into per-parameter recommendations.
Yet such guides capture only \emph{what} experts conclude, discarding \emph{how} they reason.
This fundamental gap manifests in three concrete deficiencies:
documentation grows stale as software evolves, fails under heterogeneous workloads, and ignores inter-parameter dependencies.

We propose shifting from \emph{static documentation} to \emph{dynamic action} for system tuning.
We introduce \name{}, which translates expert tuning methodologies into executable
skills that equip LLM-based agents to perform version-consistency verification,
workload-specific profiling, and multi-parameter joint optimization. Evaluated
on PostgreSQL under TPC-C and TPC-H benchmarks, \name{} outperforms
state-of-the-art documentation-driven tuning baselines by up to 35.2\%. 
The tool is available at \url{https://github.com/ISCAS-OSLab/PerfEvolve}.

\end{abstract}

\section{Introduction}

Modern computer systems, from operating systems~\cite{windows, linux, freebsd} to database systems~\cite{pg, mysql8, redis, rocksdb}, expose hundreds of configurable parameters that govern their runtime behavior. These parameters control everything from memory allocation to I/O scheduling to concurrency policies. Collectively, they define a vast configuration space whose optimal point shifts with hardware generations, workload characteristics, and 
deployment scale. Navigating this space well can yield order-of-magnitude 
performance improvements~\cite{challenge, selfdriving, sheep, best}; navigating it 
poorly leads to resource waste, latency spikes, and missed service-level objectives.

Historically, system operators have relied on \emph{documentation} as the primary 
source for tuning expertise: official reference manuals, community-maintained 
guides, and the built-in rule sets embedded in dedicated tuning 
tools~\cite{ottertune, cdbtune, ituned, pg-tuning, mysql-tuing, redis-tuing, rocksdb-tuning}.
Documentation encodes the expertise of system developers and practitioners, 
distilling years of benchmarking and field observation into recommended defaults, safe operating ranges, and cautionary notes.
In a sense, it is the externalized long-term knowledge of the systems community~\cite{sre-book, debug, blackwhite}.

The advent of large language models (LLMs) has opened a new frontier in 
automated tuning.
Recent work has proposed feeding documentation 
directly to LLMs to construct informed priors over the parameter space~\cite{dbbert, lambda, rabbit}.
State-of-the-art systems such as GPTuner~\cite{gptuner} exemplifies this paradigm: it ingests official manuals 
and community guides, uses LLMs to extract parameter semantics and value 
constraints, and injects the resulting knowledge into a Bayesian optimization loop~\cite{facilitating}, which significantly narrows the search space compared to uninformed baselines.
However, this line of work rests on an implicit assumption that 
\emph{documentation is sufficiently correct}. When this assumption is flawed, improvements are bounded by the quality of the documentation itself.

Through a systematic analysis of documentation for various production
systems~\cite{pg, mysql8, rocksdb, linux, kafka}, we identify three deficiencies that hinder documentation's utility as a tuning oracle.
To ground our analysis, we select a widely deployed relational database system, i.e., PostgreSQL, as a representative running example.
PostgreSQL offers a mature documentation ecosystem and has served as the primary benchmark for recent documentation-driven tuning frameworks~\cite{gptuner, e2etune}.
\begin{myitemize}
    \item \emph{Staleness:} Documentation lags behind system evolution. PostgreSQL's \texttt{random\_page\_cost} still recommends~4.0 since version 7.4, calibrated for rotational HDD, yet the optimal value on SSDs is closer to~1.1~\cite{conf, pgtune}.
    \item \emph{Context insensitivity:} Documentation specifies \emph{what} a parameter does but not \emph{when} a given value is optimal. PostgreSQL recommends \texttt{shared\_buffers}~$=$~25\% RAM with no distinction between OLTP and OLAP, SSD and HDD, or read- vs.\ write-heavy workloads~\cite{index, oracle9i, agentune}.
    \item \emph{Correlation absence:}
    Documentation overlooks parameter correlation.
    Our analysis shows that the \texttt{shared\_buffers} $\times$ \texttt{work\_mem} correlation explains 39\% of performance variance, a dependency entirely absent from the official documentation.
\end{myitemize}

\heading{Insight: process (\emph{how}), rather than results (\emph{what}).}
The deficiencies above share a common root cause. Documentation records the 
\emph{results} of expert tuning, the recommended values that practitioners 
have concluded through experience, rather than the \emph{process} by which 
experts arrive at those recommendations.
A result
(e.g., ``set \texttt{shared\_buffers} to 25\% of RAM'')
is an 
environment-specific snapshot. It embeds implicit assumptions about hardware, 
workload, and system version at the time of writing. When any of those 
assumptions shift, a new server is provisioned, a workload evolves, a version 
upgrade changes internal behavior, the recommendation becomes stale or 
misleading.
A process (e.g., ``measure buffer hit ratio under representative load; adjust 
\texttt{shared\_buffers} across candidate values") is 
environment-agnostic.
It is a reproducible methodology that seeks the best answer for the target environment and workload it is applied to. The process does not become 
stale when the system version changes.

This distinction has a practical implication for automated tuning. In any 
deployment scenario, active \emph{profiling} of the 
target system under a representative workload is necessary to determine the 
true optimal configuration. Documentation should serve to \emph{accelerate} 
profiling: telling the tuning agent what to measure, how to structure its probes, 
and what decision rules to apply to the signals it observes. Documentation that 
only provides static recommended values offers none of this guidance; it either 
leads the agent to blindly apply stale recommendations or reduces it to 
uninformed random search over the huge configuration space.

\heading{Proposal.}
We present \name, a tool that operationalizes expert tuning 
methodology as \emph{executable procedural knowledge}. Rather than encoding 
tuning expertise as value or range recommendations, \name encodes it as structured procedural knowledge that an LLM-based tuning agent can learn and execute against a live system to determine the optimal configuration for that specific deployment. \name accelerates profiling through two novel techniques:
\begin{myitemize}
    \item \emph{Dimensionality reduction via sensitivity analysis:}
    From hundreds of parameters, \name identifies the small subset that 
    materially affects performance for the given workload via top-$k$ sensitivity 
    analysis. Focusing profiling effort on this subset reduces the effective search 
    space by orders of magnitude and concentrates optimization budget on the 
    parameters where it yields the greatest reward.
    \item \emph{Topology discovery for joint optimization:}
    Before committing to any configuration, \name mines the correlation 
    structure among sensitive parameters and groups strongly correlated ones 
    into joint optimization components.
    Parameters within a component are tuned together, capturing correlation effects that per-parameter tuning would miss.
\end{myitemize}

We compare \name against two state-of-the-art systems, GPTuner~\cite{gptuner} and E2ETune~\cite{e2etune}. \name improves PostgreSQL performance by up to 35.2\% across OLTP and OLAP workloads in just 30 trials. Under cross-hardware transfer, prior approaches degrade substantially, while \name effectively recovers from these degradations by up to 58.9\% without requiring hardware-specific retuning. 
Moreover, PERFEVOLVE eliminates invalid or harmful configurations,
raising the success rate of tuning experiments from 68\% to 100\% and further accelerating convergence.

\heading{Contributions.}
We make the following contributions:

\begin{myitemize}
    \item We systematically identify three fundamental deficiencies that hinder existing system documentation from serving as an effective and actionable tuning oracle.
    \item We propose a methodology that encodes \emph{"procedural knowledge"} into system documentation, transforming static recommendations (what to set) into reproducible workflows (how to determine what to set).
    \item We design and implement \name, a tool that automates the generation of process-oriented tuning documentation. \name provides structured skills that empower LLM-based agents to conduct autonomous profiling.
    \item \name significantly improves the system performance over state-of-the-art tuning systems.
\end{myitemize}
\section{Characterization of Documentation Gaps}
\label{sec:gap}


\subsection{Documentation Deficiencies Across Systems}
\label{sec:gap:cross}

We survey the official documentation of four widely-deployed systems, including Linux~\cite{linux}, PostgreSQL~\cite{pg}, MySQL~\cite{mysql8}, and Apache Kafka~\cite{kafka}, and identify three types of recurring structural deficiencies.
Table~\ref{tab:cross-system} summarizes representative examples; we elaborate on each deficiency below.

\begin{table*}[t]
\centering
\caption{General deficiencies in systems tuning documentation. Each entry contrasts what official documentation provides with what is missing for effective tuning, and summarizes empirically observed consequences (when available).}
\label{tab:cross-system}
\footnotesize
\setlength{\tabcolsep}{3pt}
\begin{tabular}{|>{\centering\arraybackslash}m{3.0cm}|
                >{\centering\arraybackslash}m{2.5cm}|
                >{\centering\arraybackslash}m{3.0cm}|
                >{\centering\arraybackslash}m{4.5cm}|}
\hline
\textbf{System Parameter} & \textbf{What documentation provides} & \textbf{What is missing} & \textbf{Observed impact} \\
\hline

\multicolumn{4}{|c|}{\textbf{Staleness (outdated operational assumptions)}} \\
\hline

Linux \texttt{vm.dirty\_ratio}
& Default (20\%) designed for older storage with higher write latency.
& Lack of updated guidance for modern SSD flush behavior.
& High values delay writeback, causing bursty I/O and latency spikes under write-heavy DB workloads. \\
\hline

Linux \texttt{vm.swappiness}
& Default value (60) with general guidance on swap aggressiveness.
& Lack of reflecting modern memory reclamation (e.g., Multi-Gen LRU) and modern swap (e.g., Zram).
& Suboptimal memory balance under modern workloads; may degrade performance under memory pressure. \\
\hline

PostgreSQL \texttt{random\_page\_cost}
& Default value (4.0) with qualitative guidance assuming HDD-era storage.
& No hardware-aware recalibration methodology for modern SSD/NVMe deployments.
& On SSD-backed workloads, optimal values depend on workload type: lowering to 1.0 improves write-heavy OLTP by 1.4$\times$ but degrades read-heavy OLTP by 24\%. \\
\hline

PostgreSQL \texttt{effective\_cache\_size}
& Default value (4GB)
& No guidance for dynamic environments where cache availability fluctuates.
& Misestimation leads to suboptimal plan selection (e.g., index scans avoided unnecessarily). \\
\hline

\hline

\multicolumn{4}{|c|}{\textbf{Context Insensitivity (generic defaults without workload conditioning)}} \\
\hline

PostgreSQL \texttt{shared\_buffers}
& Heuristic recommends 25\% of RAM as a reasonable starting point on a dedicated server.
& No adaptation for workload type (OLTP vs OLAP), storage medium, and read/write ratio.
& Optimal values vary across workloads; misconfiguration leads to 5--16\% throughput degradation on SSD-heavy write workloads. \\
\hline

MySQL \texttt{innodb\_buffer\_pool\_} \texttt{size}
& Only description of the legality
& No consideration of co-located services and memory contention.
& May trigger memory contention under co-tenancy, potentially leading to OOM events and unstable performance. \\
\hline

Kafka \texttt{num.partitions}
& Suggested based on expected throughput or consumer count.
& Lack of consideration of partitioning numbers, consumer parallelism, and workload skew.
& Misalignment leads to limited scalability and suboptimal throughput due to load imbalance. \\
\hline

\multicolumn{4}{|c|}{\textbf{Correlation Absence‌ (missing multi-parameter modeling)}} \\
\hline

PostgreSQL \texttt{shared\_buffers} $\times$ \texttt{work\_mem}
& Parameters are documented largely independently.
& No correlation-aware guidance on joint tuning.
& Our ANOVA shows that 
tuning parameters independently led to measurable throughput degradation. \\
\hline

PostgreSQL \texttt{work\_mem} $\times$ \texttt{max\_connections}
& Both parameters documented separately.
& No explicit coupling constraint on total memory consumption.
& High connection counts amplify per-query memory usage, causing OOM or instability. \\
\hline

Kafka partitions $\times$ consumers
& Separate guidelines for partition count and consumer threads.
& No cross-component coordination strategy.
& Throughput plateaus when partitioning is misaligned with consumer parallelism. \\
\hline

\end{tabular}

\end{table*}

\heading{Staleness.}
Documentation often encodes tuning recommendations as constants calibrated to the hardware and operating systems of their time.
Yet modern systems evolve continuously, from CPU and GPU architectures to high-performance NVMe storage and RDMA networking, causing these defaults to silently become stale.
For example, in Linux memory management, \texttt{vm.swappiness}\,=\,60 predates modern reclamation techniques (e.g., multi-gen LRU) and modern swap (e.g., Zram). \texttt{vm.dirty\_ratio}=20\% was calibrated for HDDs; on modern SSDs it delays writeback excessively, producing bursty I/O.
Similarly, our experiments show that on SSD-backed VMs, lowering \rpc{} to 1.0 improves write-heavy OLTP throughput by 1.4$\times$, yet \emph{degrades} read-heavy OLTP by 24\%~\ref{tab:cross-system}, because the optimal value is workload-dependent and non-monotonic (\S\ref{sec:gap:quant}).
In all cases, documentation defaults are written once and rarely revisited, while systems (both hardware and software) keep updating.

\heading{Context insensitivity.}
Documentation typically provides default values or ranges without conditioning on workload characteristics or deployment context. Yet configuration impact is inherently context-dependent: the same value can be optimal in one setting and harmful in another.
Consider PostgreSQL's canonical recommendation of 
\texttt{shared\_buffers}\,=\,25\% of RAM, this single figure elides workload type, storage medium, and read/write ratio.
In practice, misconfiguration along this single parameter alone leads to 5--16\% throughput loss
(\S\ref{sec:gap:quant}).
Similar issues appear in MySQL (\texttt{innodb\_buffer\_pool\_size}, ignoring co-tenancy and 
competing memory pressure) and Kafka (\texttt{num.partitions}, ignoring coordination with consumers), leading to instability or suboptimal throughput.
The root cause is that it is impossible for system developers to enumerate the combinatorial space of workloads, deployment environments at documentation time.

\heading{Correlation absence‌.}
In most cases, documentation describes parameters in isolation, implicitly assuming independent tuning.
However, parameter correlations are both common and often dominant.
Our experiments (\S\ref{sec:gap:interaction}) show that 60\% of parameter pairs exhibit interaction strength $>$15\%.
In particular, the interaction between \texttt{shared\_buffers} and \texttt{work\_mem} explains 39\% of performance variance, measured by the ANOVA effect size $\eta^2$ (i.e., the fraction of total variance attributable to the correlation term).
Critically, many strong correlations span subsystems (e.g., memory $\times$ concurrency), yet remain undocumented.

\subsection{Case Study: PostgreSQL with 300+ Parameters}
\label{sec:gap:quant}

We select PostgreSQL as our case study for three reasons.
First, it exposes $300$+ tunable parameters spanning I/O, memory, concurrency, query planning, etc.
Second, it has a mature documentation ecosystem with extensive best-practice guidelines from practitioners.
Third, it is the target of various documentation-driven tuning approaches~\cite{gptuner,e2etune,L2T-Tune}, making it the ideal system for evaluating documentation quality.

We deploy PostgreSQL v16 instances on 150 virtual machines with identical hardware configurations (8\,vCPU, 8\,GB DRAM, 200\,GB SSD).
We evaluate three representative workloads: \textbf{TPC-C-r} (read-heavy OLTP), \textbf{TPC-C-w} (write-heavy OLTP), and \textbf{TPC-H} (analytical OLAP).
All results report performance change relative to the default PostgreSQL configuration; each data point is averaged over three independent runs.

\heading{Finding-\#1: Established best practices may become obsolete.}
\label{sec:gap:staleness}
We measure two widely-used baselines: PostgreSQL official guidelines (PG-Official) and PGTune~\cite{pgtune}, which translates documentation into heuristic, resource-aware rules but lacks workload adaptivity and correlation awareness. 

\begin{figure}[htbp]
  \centering
  \includegraphics[width=0.64\columnwidth]{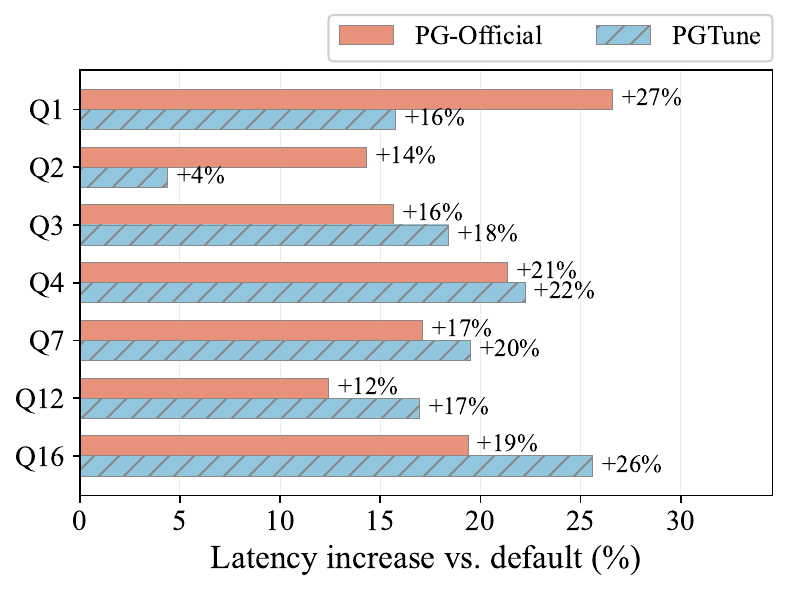}
  \caption{Latency increase on TPC-H when applying PG-Official and PGTune rules (7 of 22 queries degraded by $>$10\%).
  Both rule sets lead to worse latency on the same kinds of sort- and aggregation-intensive queries.
  }
  \label{fig:tpch-perquery}
\end{figure}

Figure~\ref{fig:tpch-perquery} drills into TPC-H's 22 queries and pinpoints 7 out of them that both rule sets degrade by more than 10\%.
The affected queries share a common trait: they are sort- or aggregation-intensive (e.g., Q1 performs a full-table aggregation with sorting; Q7 joins multiple tables with an order-by clause).
The root cause is a memory allocation mismatch.
Both PG-Official and PGTune set \texttt{shared\_buffers\,=\,2\,GB} (the canonical ``25\% of RAM'' rule), which shrinks the OS page cache available for temporary files, while leaving \texttt{work\_mem} at its 4\,MB default, which is too small for the large intermediate results these queries produce.
As a result, sort and hash operations spill to disk, inflating latency by 12--27\%.

A second example reinforces the pattern.
Both rule sets recommend \texttt{\cct{}\,=\,0.9} (``spread checkpoint writes over 90\% of the interval''), advice calibrated for rotational disks where I/O bursts are expensive.
On our SSD-backed VMs, this recommendation \emph{hurts}: a single-parameter sweep across 6{,}297 configurations shows that the default value (0.5) outperforms 0.9 by 15\% on read-heavy OLTP, and the optimal value (0.0, i.e., checkpoint as fast as possible) outperforms 0.9 by 24\% on write-heavy OLTP.

These failures share a common cause: documentation records \emph{conclusions} derived under specific assumptions, but provides no mechanism to detect when those assumptions are violated.


\begin{figure}[htbp]
  \centering
  \includegraphics[width=0.64\columnwidth]{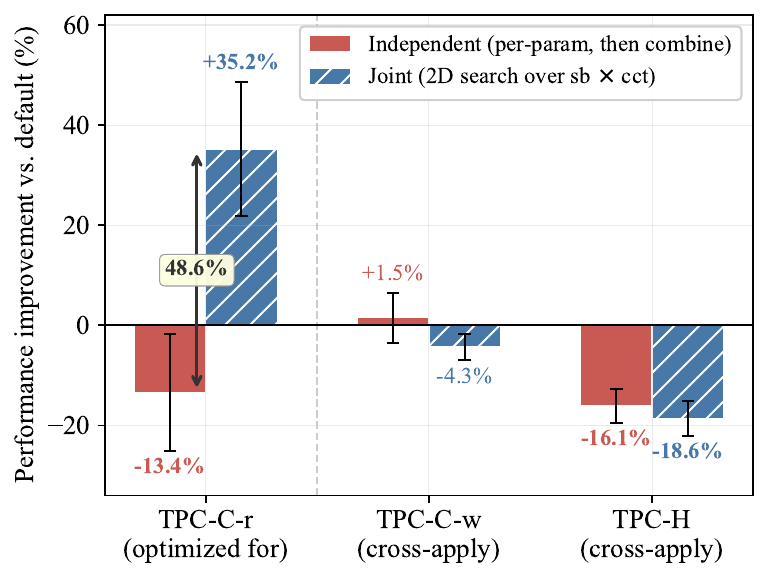}
  \caption{Independent vs.\ joint optimization for \texttt{shared\_buffers} (sb) $\times$ \texttt{checkpoint\_completion\_target} (cct) across three workloads. ``Independent'' tunes each parameter to its individually-optimal value in isolation, then combines the individually optimal values. ``Joint'' searches the two-dimensional space together.}
  \label{fig:indep-vs-joint}
\end{figure}

\heading{Finding-\#2: Optimal configurations are workload-dependent.}
\label{sec:gap:workload}
System documentation typically provides generic, one-size-fits-all recommendations.
In practice, however, the effectiveness of a configuration is heavily conditioned on workload characteristics, including read/write ratio, access patterns, and resource bottlenecks.

Our study shows that most documentation typically specifies a single heuristic or default value per parameter.
For example, \texttt{shared\_buffers} is commonly recommended as a fixed fraction of memory (e.g., 25\%), and checkpoint-related parameters such as \texttt{checkpoint\_completion\_target} are described in terms of general trade-offs (e.g., smoothing I/O).
These recommendations are presented as broadly applicable starting points.

Such descriptions do not capture how parameter choices should adapt to workload-specific conditions.
They lack guidance on when a configuration that benefits one workload (e.g., read-heavy OLTP) should be adjusted or even reversed for another (e.g., write-heavy OLTP or OLAP).
In particular, documentation does not specify how changing bottlenecks (buffer reuse vs.\ logging vs.\ large scans) alter the role of each parameter.

We observe that configurations optimized for one workload often fail to transfer. As shown in Figure~\ref{fig:indep-vs-joint}, a configuration that improves throughput by \textbf{+35.2\%} on TPC-C-r degrades performance by \textbf{$-$4.3\%} on TPC-C-w and by up to \textbf{$-$18.6\%} on TPC-H.
Even configurations obtained from careful per-parameter tuning exhibit similar issues when applied across workloads.
Overall, the performance gap between workload-specific optima can reach up to \textbf{48.6\%}, indicating that misaligned configurations can significantly harm system performance.

These results highlight a fundamental limitation: there is no universally optimal configuration.
Effective tuning requires workload-aware adaptation, yet existing documentation cannot indicate when or how configurations should change.

\begin{figure}[t]
  \centering
  \includegraphics[width=0.84\textwidth]{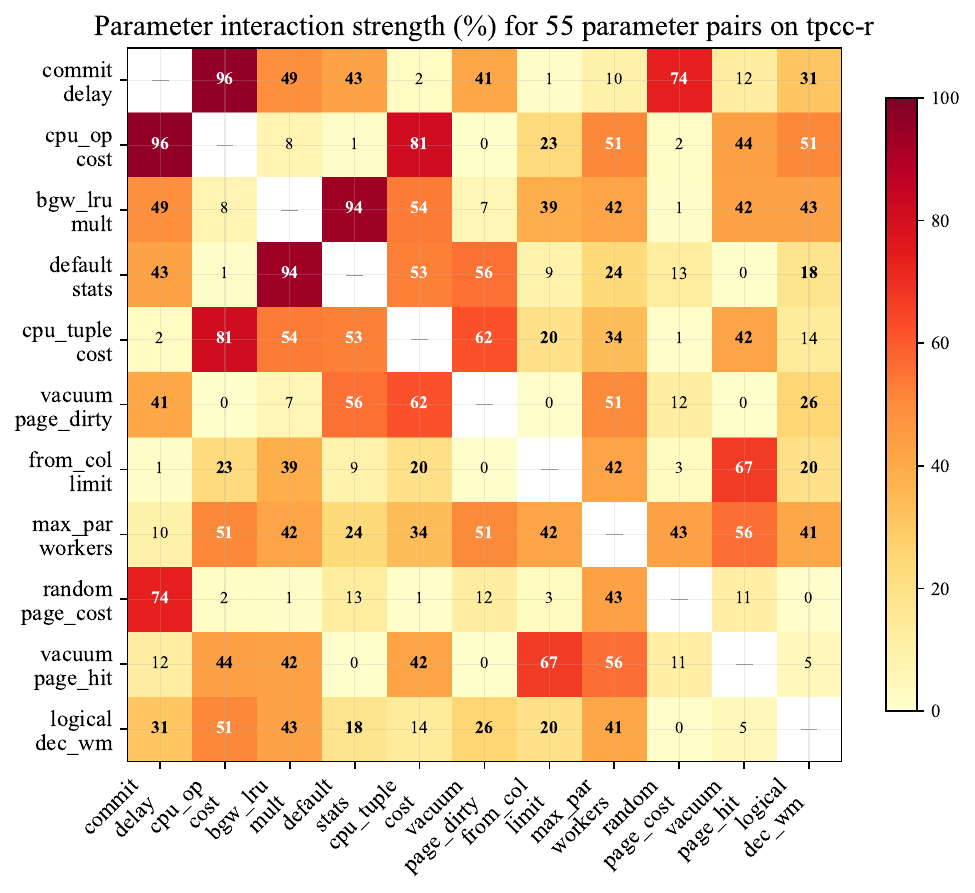}
  \caption{Interaction strength for 55 parameter pairs on TPC-C-r. Each cell shows the interaction percentage; cells above 15\% (bold) indicate candidate correlations for fine-grained verification. 33 of 55 pairs (60\%) exceed the threshold.}
  \label{fig:heatmap}
\end{figure}

\heading{Finding-\#3: Parameter correlations are pervasive and can be cross-subsystem.}
\label{sec:gap:interaction}
To determine which parameters should be jointly tuned, we quantify pairwise correlations using the two-stage methodology in \S\ref{sec:design:topo}.
We select 11 high-sensitivity parameters and exhaustively test all $\binom{11}{2} = 55$ pairs via $2 \times 2$ factorial experiments ($n = 1$) on TPC-C-r.
Table~\ref{tab:top-interactions} ranks the strongest correlation.

\begin{table}[htbp]
\centering
\caption{Top-10 parameter interactions. ``Int.\%'' is the interaction strength. 8 of 10 span different PostgreSQL subsystems.}
\label{tab:top-interactions}
\small
\begin{tabularx}{\columnwidth}{%
  >{\raggedright\arraybackslash}X
  >{\raggedright\arraybackslash}X
  >{\raggedleft\arraybackslash}X
  >{\centering\arraybackslash}X
}
\toprule
\textbf{Param.\ $A$} & \textbf{Param.\ $B$} & \textbf{Int.\%} & \textbf{Cross?} \\
\midrule
commit\_delay       & cpu\_op\_cost       & \textbf{96.0} & \ding{51} \\
bgw\_lru\_mult      & default\_stats      & \textbf{93.7} & \ding{51} \\
cpu\_op\_cost       & cpu\_tuple\_cost    & 80.9 & \ding{55} \\
commit\_delay       & random\_page\_cost  & 73.6 & \ding{51} \\
from\_col\_limit    & vac\_page\_hit      & 66.8 & \ding{51} \\
cpu\_tuple\_cost    & vac\_page\_dirty    & 62.0 & \ding{51} \\
max\_par\_workers   & vac\_page\_hit      & 56.0 & \ding{51} \\
default\_stats      & vac\_page\_dirty    & 55.5 & \ding{51} \\
bgw\_lru\_mult      & cpu\_tuple\_cost    & 53.5 & \ding{51} \\
cpu\_tuple\_cost    & default\_stats      & 52.6 & \ding{55} \\
\bottomrule
\end{tabularx}
\end{table}

Figure~\ref{fig:heatmap} shows that \textbf{33 of 55 pairs (60\%) exceed the 15\% correlation threshold}, indicating that parameter correlations are the norm rather than exceptions. The strongest pair, \texttt{commit\_delay} $\times$ \texttt{cpu\_operator\_cost}, reaches 96\% interaction strength (Int.\%), measured as the fraction of performance variance, indicating that the effect of one parameter largely depends on the other.
Moreover, correlations are frequently cross-subsystem: 8 of the top-10 pairs span different PostgreSQL components (e.g., WAL $\times$ planner, background writer $\times$ query optimizer), yet such relationships are absent from documentation.

\heading{Summary.}
\label{sec:gap:summary}
These three findings share a common root cause: documentation encodes \emph{what to set}, but not \emph{how to determine what to set}. 
Today's documentation tuning recommendations are tied to specific hardware (staleness), blind to workload context (insensitivity), and isolated per parameter (independence).
This motivates us to seek a better form of knowledge that encodes the \emph{process} of tuning, instead of the results.

\section{\name{} Design}
\label{sec:design}

At a high level, \name{} is a tool that serves as an automated engine which synthesizes procedural tuning documents from a comprehensive profiling process. \name{}'s ``step-by-step'' design philosophy allows it to leverage the reasoning capabilities of modern LLM-based agents~\cite{is, react, deps, reflexion, reasoning, agent}, thereby accelerating the profiling process on diverse deployment environments.
Specifically, \name{} introduces two novel techniques: \emph{dimensionality
reduction} and \emph{topology discovery}, which together make profiling practical at scale. Dimensionality reduction identifies the small subset of parameters that materially affect performance; topology discovery reveals which of those parameters are correlated and must therefore be co-optimized. 

\begin{figure}[h]
  \centering
  \includegraphics[width=0.76\columnwidth]{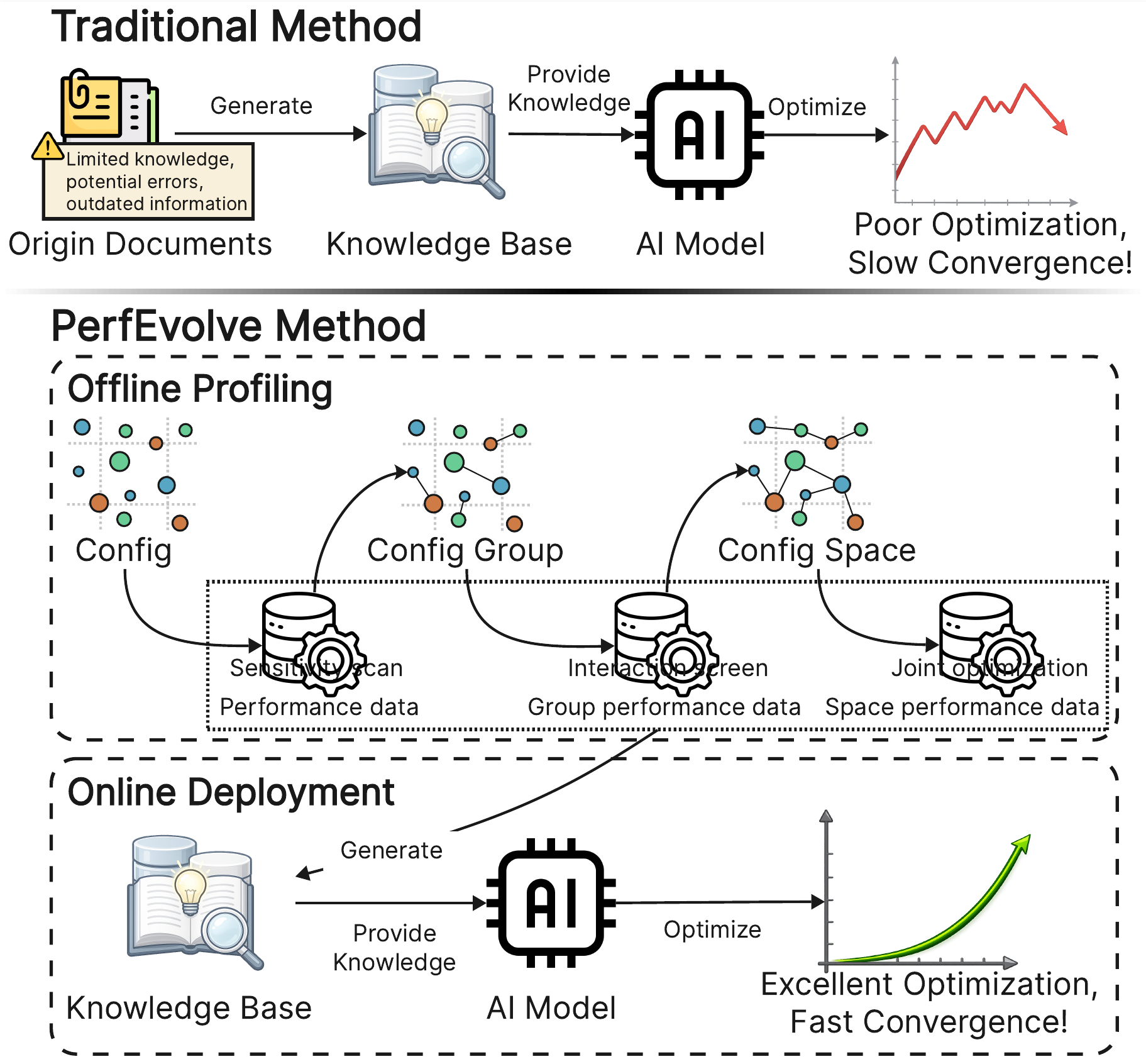}
  \caption{\name{} architecture. The \emph{offline profiling phase} runs once per system version: it applies dimensionality reduction and topology discovery to produce a procedural tuning document. The \emph{online deployment phase} runs per deployment: an LLM agent executes the document's skills, performing targeted profiling guided by the offline-derived reference data.}
  \label{fig:arch}
\end{figure}


\heading{Goals.}
\name{} is designed around three objectives:
\begin{myitemize}
\item \emph{Executable:}
The documentation should define concrete actions (e.g., ``run benchmark $B$ with configuration $C$'', ``compute $\eta^2$ from results'') and
explicit decision rules, enabling reproducible step-by-step execution by an agent (either human-based or LLM-based) without ambiguity.

\item \emph{Verifiable:}
Post-conditions (predicates expected to hold after execution) and procedural knowledge (relationship from offline profiling to support decisions) should support self-checking at each
step. For example, if $\leq$ 5 sensitive parameters are
found (violating ``$|\mathit{top\text{-}k}| \geq 5$''), the
agent detects the anomaly and triggers an adaptation skill.

\item \emph{Transferable:}
Procedures and decision rules are environment-agnostic. Adapting to a new version
requires updating calibration data, not rewriting the document.
\end{myitemize}

\heading{Workflow.} \name's workflow consists of two phases: an offline phase (run once per software version), and an online phase (run per deployment).
During the offline profiling phase, given a target system with $n$ tunable parameters and $W$
representative workloads, \name{} executes the two techniques in
sequence:
\begin{myitemize}
\item \emph{Dimensionality reduction}: scan all $n$ parameters to
identify the $k \ll n$ that significantly affect performance, along
with their safe operating ranges and response-curve shapes.
\item \emph{Topology discovery}: screen all $\binom{k}{2}$
parameter pairs for correlation, construct a graph,
and decompose it into connected components.
\end{myitemize}
The outputs are fed into a \emph{document generator} that produces a
set of executable skills (\S\ref{sec:design:docgen}).
During the online deployment phase, an LLM-based tuning agent receives the procedural document and
executes its skills against the target deployment. The document tells
the agent \emph{what} to measure and \emph{how} to interpret the
results; the agent performs only the targeted experiments prescribed
by each skill. Because the offline phase has already identified which
parameters matter (via dimensionality reduction) and which must be
co-optimized (via topology discovery), the online phase converges in
tens of experiments rather than thousands.
Figure~\ref{fig:arch} illustrates the two-phase architecture.

\subsection{Dimensionality Reduction}
\label{sec:design:dim}

\heading{Problem statement.}
A complex system may expose hundreds of tunable parameters, yet
exhaustive exploration of all of them is intractable~\cite{toomany, expand}. Worse, existing
documentation provides no principled guidance on which parameters have
meaningful performance impact, leaving practitioners to tune
blindly.

\heading{Top-$k$ sensitivity analysis.}
We perform a single-parameter sweep for each of the $n$ parameters:
for parameter $x_i$, we fix all others at their defaults and evaluate
$L_i \in \{3, \ldots, 9\}$ values spread across the documented or
empirically safe range. Each configuration is benchmarked with $r=3$
repetitions under each of the $W$ workloads.
For each parameter $x_i$ and workload $w$, we compute the
\emph{coefficient of variation}~\cite{cv}:
\begin{equation}
  \text{CV}_{i,w} = \frac{\max_v \overline{\text{TPS}}(x_i{=}v,\,w)
    - \min_v \overline{\text{TPS}}(x_i{=}v,\,w)}
    {\overline{\text{TPS}}(x_i{=}\text{default},\,w)}
  \label{eq:cv}
\end{equation}
where $\overline{\text{TPS}}$ denotes the mean over repetitions. We
define the aggregate sensitivity as $\text{CV}_i = \max_w
\text{CV}_{i,w}$, and select the top-$k$ parameters exceeding a
threshold $\tau_s$ (5\% by default).
The total sweep requires $\sum_{i=1}^n L_i \times r \times W$
experiments. For PostgreSQL's 116 performance-relevant parameters, this amounts to
$116 \times \bar{L} \times 3 \times 3 \approx 6{,}297$ benchmark
runs. Because each parameter's sweep is independent of all others,
this phase is \emph{embarrassingly parallel}: on our 150-server
cluster, it completes within one day.

\heading{Takeaway.}
Dimensionality reduction produces 4 artifacts in the procedural document:
\begin{myitemize}
  \item \emph{Sensitivity ranking}: parameters ordered by
  $\text{CV}_i$. In our study, only 15 of 116 parameters exceed the
  5\% threshold; the remaining ${\sim}100$ have $\text{CV} < 1\%$
  and can safely remain at their defaults.

  \item \emph{Safe ranges}: for each top-$k$ parameter, the
  empirically observed range $[\min_v, \max_v]$ within which no
  crash or severe degradation ($>$50\% throughput loss) occurred.
  These ranges eliminate the unsafe configurations responsible for
  the 32\% crash rate observed in GPTuner.

  \item \emph{Response-curve shapes}: each parameter is classified
  as \emph{monotonic-up}, \emph{monotonic-down}, \emph{non-monotonic}
  (interior optimum), or \emph{step-function} (threshold effect).
  This classification directly informs the agent's search strategy:
  monotonic parameters require only boundary evaluation, whereas
  non-monotonic ones require finer sweeps around the peak.

  \item \emph{Per-workload sensitivity}: different workloads may
  induce different top-$k$ sets~\cite{agentune}. Recording workload-specific rankings
  allows the agent to adapt its focus when the deployment workload
  diverges from the offline representative.
\end{myitemize}

\begin{figure}[t]
  \centering
  \includegraphics[width=0.76\columnwidth]{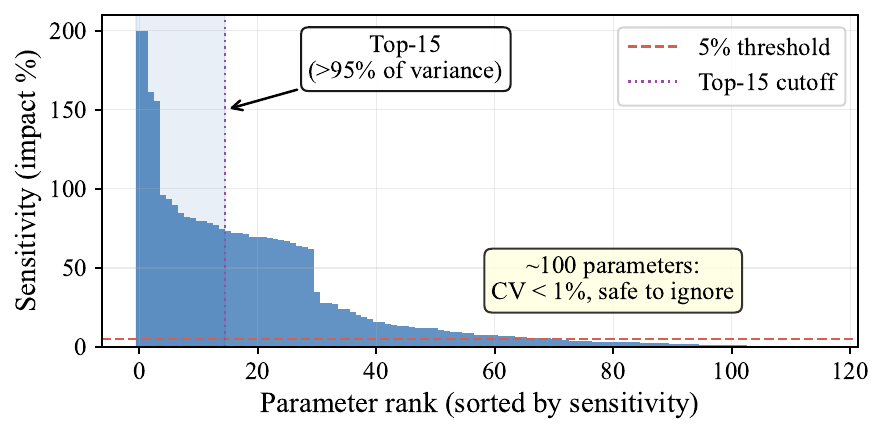}
  \caption{Sensitivity ($\text{CV}$) of all 116 PostgreSQL performance-relevant parameters sorted in descending order, on the \texttt{TPC-C-r} workload. The
  distribution exhibits a pronounced long tail: the top 15 parameters
  account for $>$95\% of cumulative performance variation. The dashed
  line marks the 5\% threshold $\tau_s$.}
  \label{fig:longtail}
\end{figure}

Figure~\ref{fig:longtail} shows the resulting sensitivity
distribution. The long-tail structure, a handful of high-impact
parameters and a large mass of negligible ones, is precisely the
property that makes dimensionality reduction effective. By narrowing
the search space from $n=116$ to $k=15$, we achieve a
\textbf{one-order-of-magnitude reduction} in the parameters that
downstream optimization must explore.

\subsection{Correlation Topology Discovery}
\label{sec:design:topo}

\heading{Problem statement.}
Reducing to $k$ parameters is necessary but not sufficient: if
parameters have implicit correlation, tuning them independently can be actively harmful. Existing documentation has little knowledge on
parameter correlations, leaving optimizers to assume independence by
default. We aim to discover the \emph{
topology}, determining which parameters must be jointly optimized and
which can safely be treated in isolation.

\heading{Two-stage factorial analysis of variance (ANOVA)~\cite{cohen1988}.}
We screen all $\binom{k}{2}$ parameter pairs through a two-stage
process designed to balance statistical rigor with experimental cost.

\emph{Stage A: Coarse screening.}
For each pair $(x_i, x_j)$, we run a $2 \times 2$ factorial
experiment (4 configurations, 1 repetition per workload), using the
extreme values from the sensitivity scan as the two factor levels. We
compute an approximate interaction percentage:
\begin{equation}
  \text{Int\%}_{ij} = \frac{|\overline{y}_{++} - \overline{y}_{+-}
    - \overline{y}_{-+} + \overline{y}_{--}|}
    {|\overline{y}_{++} + \overline{y}_{+-} + \overline{y}_{-+}
    + \overline{y}_{--}|/4}
  \label{eq:coarse}
\end{equation}
where $\overline{y}_{ab}$ is the mean TPS at level $(a, b)$. Pairs
with $\text{Int\%} > 15\%$ advance to Stage~B; pairs below 5\% are
marked independent. This threshold is intentionally liberal: the
$2 \times 2$ design with $n=1$ overestimates interaction strength by
approximately $2\times$ (validated experimentally), so Stage~A
maximizes recall at the cost of some false positives.


\emph{Stage B: Fine screening.}
For each pair passing Stage~A, we run a $4 \times 4$ factorial design
with $r=3$ repetitions per cell (48 runs per pair per workload) and measure interaction strength using effect size (partial $\eta^2$) following standard ANOVA practice~\cite{lakens2013effect}. We report the partial
effect size:
\begin{equation}
  \eta^2_{ij} = \frac{\text{SS}_{ij}}{\text{SS}_{\text{total}}}
  \label{eq:eta2}
\end{equation}
which captures the proportion of variance explained by the correlation term. We apply Benjamini-Hochberg FDR correction \cite{fdr} across all tested pairs
and retain those with $\eta^2 > 0.15$ and corrected $p < 0.05$ as
confirmed correlations.

\heading{Building correlation graphs.}
We construct a weighted undirected graph $G = (V, E)$ where $V$ is
the set of $k$ sensitive parameters and each edge $(i,j) \in E$
carries weight $\eta^2_{ij}$ for every confirmed correlation. We then
decompose $G$ into connected components $\{C_1, C_2, \ldots, C_m\}$.
Parameters within the same component \emph{must} be jointly
optimized; parameters in different components or isolated
nodes can be tuned independently without loss of optimality.

\begin{figure*}[t]
  \centering
  \includegraphics[width=\textwidth]{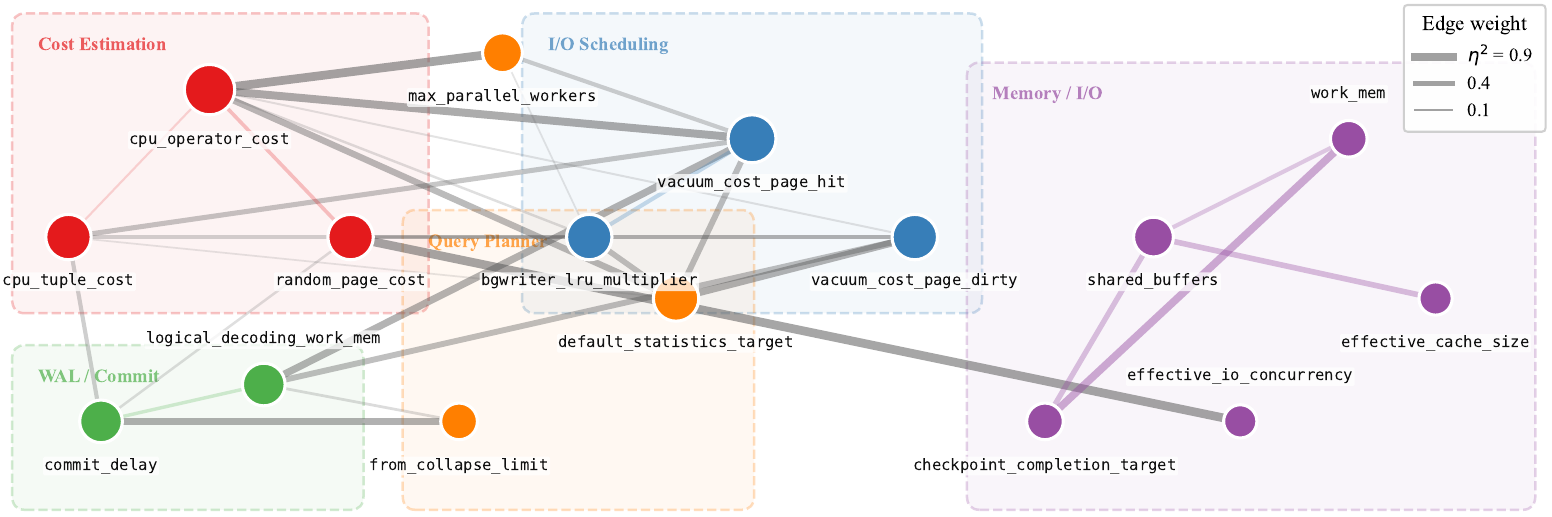}
  \caption{Correlation graph for PostgreSQL on the \texttt{TPC-C-r}
  workload. \textbf{Left}: nodes are the top-15 parameters; edges
  indicate confirmed correlations ($\eta^2 > 0.15$, $p < 0.05$);
  edge thickness is proportional to $\eta^2$; colors denote connected
  components. \textbf{Right}: the resulting optimization strategy.
  Component $C_1$ (\texttt{shared\_buffers}, \texttt{work\_mem},
  \texttt{checkpoint\_completion\_target}) requires a 3-dimensional
  joint search; $C_2$ requires 2-dimensional; isolated nodes are
  tuned independently.}
  \label{fig:interaction-graph}
\end{figure*}

\heading{Takeaway.} Interestingly, we observe a structural sparsity property that makes component-wise optimization computationally tractable in practice.
Although 64\% of PostgreSQL parameter pairs exhibit statistically
significant correlation, the graph decomposes into low-dimensional components.
Specifically, the largest component contains only 3–4 parameters.
This decomposition reduces the search space from an intractable $4^{15}$ global configurations to a series of manageable sub-problems (e.g., $4^3 \times 3 = 192$ runs for a 3-parameter component), making exhaustive local optimization feasible.


\subsection{Procedural Document Generation}
\label{sec:design:docgen}

We formalize a \textbf{procedural tuning document} as a directed acyclic graph (DAG) of \emph{skills}, where each skill is a self-contained executable tuning unit. Specifically, a skill is a tuple of the following elements:
\begin{myitemize}
  \item $\mathit{preconditions}$: predicates required before execution (e.g., ``baseline measurement must be completed'');
  \item $\emph{procedure}$: an ordered sequence of actions (benchmark, measure, compare) with conditional branches;
  \item $\emph{decision criteria}$: rules selecting execution paths based on signals (e.g., ``if $\eta^2 >
  0.15$, trigger joint optimization'');
  \item $\emph{postconditions}$: predicates expected after execution (e.g., ``optimal configuration identified
  with $>$95\% confidence'').
  \item $\emph{reference data}$: empirical data
  from offline profiling that anchor the decision criteria (e.g., sensitivity thresholds).
\end{myitemize}

The outputs of dimensionality reduction (\S\ref{sec:design:dim}) and
topology discovery (\S\ref{sec:design:topo}) are compiled into an executable procedural document structured as a DAG of skills. 
The generator produces three types of skills:
\begin{myitemize}
\item \emph{Per-parameter skills:} For each parameter in the top-$k$ set: a sensitivity profile (i.e, CV, response shape, safe range, workload-specific ranking) and a verification step that checks whether the documented behavior holds on the target system.

\item \emph{Per-component skills:} For each connected component in the correlation graph: a joint optimization skill specifying the search space and decision criteria for adopting joint vs.\  independent tuning.

\item \emph{Orchestration skill:} A root-level skill that orchestrates the full tuning workflow, including diagnosis, 
sensitivity scanning, correlation checking, joint optimization, and 
cross-workload verification, with each transition governed by explicit decision criteria.
\end{myitemize}
For PostgreSQL v16, the generated document contains 23 skills and 160 parameter profiles in total.

\subsection{Putting It All Together}
\label{sec:design:agent}

\name operates in one of the following two modes.

\heading{Mode 1: Full procedural execution.}
The LLM-based tuning agent receives the complete skill DAG as part of its system
prompt. At each skill, it interprets the procedure, invokes
benchmarks via a system API, evaluates decision criteria against the
observed results, and transitions to the next skill. This mode
realizes the full ``step-by-step'' paradigm: the agent
performs deployment-specific profiling guided by the
offline-derived experience, and converges without any pre-baked
configuration recommendations.

\heading{Mode 2: Knowledge injection into existing frameworks.}
For compatibility with existing tuning systems, \name{} exports a
subset of its procedural knowledge in a format compatible with the
target framework. For GPTuner~\cite{gptuner}, we export
per-parameter JSON containing: (1)~the top-$k$ parameter list
(replacing GPTuner's LLM-extracted 57-parameter list); (2)~empirical safe
ranges (replacing the documentation-derived ranges that cause 32\%
crash rates); and (3)~correlation annotations as advisory hints for
the optimizer.
This mode demonstrates that \name{}'s knowledge is not tied
to any particular agent architecture: it can improve any tuning
system that accepts structured knowledge as input.

\section{Evaluation}
\label{sec:eval}

We evaluate \name{} to answer four research questions:
\begin{myitemize}
\item RQ1: Does \name achieve better database performance than state-of-the-art approaches? (\S\ref{sec:eval:e2e})
\item RQ2: 
How does procedural knowledge outperform declarative knowledge? (\S\ref{sec:eval:form})
\item RQ3: What is the offline profiling cost?
\item RQ4: What do key techniques, dimensionality reduction and topology discovery, contribute to \name? (\S\ref{sec:eval:ablation})
\end{myitemize}

\heading{Experimental setup.}
We use two hardware configurations:
\begin{myitemize}
  \item \textbf{Virtual machine cluster} (primary): 150 PostgreSQL~16 instances (2\,vCPU, 8\,GB RAM, SSD). \name{}'s offline profiling was conducted on this cluster, so it serves as the \emph{matched} evaluation environment where profiling conditions equal deployment conditions.
  \item \textbf{High-end server}: 192-core Xeon 8575C, 1\,TB RAM, 8$\times$H100 GPUs (for ML inference used by E2ETune~\cite{e2etune}). Used to evaluate cross-hardware transfer and end-to-end comparison at scale.
\end{myitemize}

\heading{Baselines.}
Table~\ref{tab:baselines} summarizes all methods.
We compare against two state-of-the-art LLM-based tuning systems:
\textbf{GPTuner}~\cite{gptuner} extracts parameter constraints from documentation via LLM and feeds them into a SMAC Bayesian optimizer~\cite{smac} (57 parameters, 30--100 iterative trials);
\textbf{E2ETune}~\cite{e2etune} fine-tunes Mistral-7B on historical tuning data and generates a complete configuration in one inference (44 parameters).
The two differ in every dimension, iterative vs.\ one-shot, explicit extraction vs.\ implicit learning, making them complementary tests of \name{}'s generality.
For each, we evaluate both the original and a \name{}-enhanced variant (knowledge injected as SMAC constraints or prompt augmentation, with zero algorithm changes).
We also include static baselines (PG-Official and PGTune~\cite{pgtune}).

\heading{Workloads.}
We employ three representative benchmarks via BenchBase~\cite{benchbase}:
\textbf{TPC-C-r} (OLTP read-heavy, SF=50, 16 terminals, 80\% OrderStatus),
\textbf{TPC-C-w} (OLTP write-heavy, SF=20, 16 terminals, 45\% NewOrder + 43\% Payment),
\textbf{TPC-H} (OLAP, SF=1, 22 queries serial).
Each configuration is benchmarked with 3 repetitions.

\begin{table}[h]
\centering
\caption{Method comparison. \name{} knowledge is injected into GPTuner (as SMAC parameter space) and E2ETune (as prompt augmentation) without changing their algorithms.}
\label{tab:baselines}
\small
\begin{tabularx}{\columnwidth}{%
  >{\raggedright\arraybackslash}X
  >{\raggedright\arraybackslash}X
  >{\centering\arraybackslash}X
}
\toprule
\textbf{Method} & \textbf{Type} & \textbf{\# Parameters} \\
\midrule
PG-Official / PGTune & Static rules & $\sim$100 \\
GPTuner~\cite{gptuner} & LLM + Bayesian opt. & 57 \\
GPTuner+\name{} & LLM + Bayesian opt. & 15 \\
E2ETune~\cite{e2etune} & LLM one-shot & 44 \\
E2ETune+\name{} & LLM one-shot + prompt & 44 \\
\bottomrule
\end{tabularx}
\end{table}

\subsection{End-to-End Performance}
\label{sec:eval:e2e}

\begin{table*}[htbp]    
  \centering                                                                          
  \small
  \caption{End-to-end performance improvement (\%) over default PostgreSQL. \emph{Cluster} results are from the VM testbed where \name{}'s profiling was conducted (profiling =    
  deployment environment; GPTuner: 100 trials).                                
  \emph{Server} results are from server (cross-hardware transfer; GPTuner: 100 trials, E2ETune: 1 inference). \textbf{Bold}: best per workload per testbed.}                                                                                                      
  \label{tab:main-results}                                                                            
  \begin{tabularx}{\columnwidth}{%
  >{\raggedright\arraybackslash}p{0.8cm}  
  >{\raggedright\arraybackslash}p{2.4cm}  
  >{\centering\arraybackslash}X
  >{\centering\arraybackslash}X
  >{\centering\arraybackslash}X
  >{\centering\arraybackslash}X
  >{\centering\arraybackslash}X
  >{\centering\arraybackslash}X
}                                                                  
  \toprule                                                                                               
   & & \multicolumn{3}{c}{\textbf{Cluster} (2\,vCPU, 8\,GB)} & \multicolumn{3}{c}{\textbf{Server}
  (192-core, 1\,TB)} \\                                                                                  
  \cmidrule(lr){3-5} \cmidrule(lr){6-8}                                                               
  \textbf{System} & \textbf{Knowledge} & TPC-C-r & TPC-C-w & TPC-H & TPC-C-r & TPC-C-w & TPC-H \\              
  \midrule                                                                                               
  \multirow{4}{*}{\shortstack[l]{\emph{GPTuner}\\\emph{(LLM} \\\emph{\& BO}}}
    & Original {(PG docs)} & +9.0 & +16.5 & $-$3.2 & $-$54.2 & $-$56.4 & $-$54.7 \\                      
    & & \emph{\small 68\% valid} & \emph{\small 70\% valid} &\emph{\small 81\% valid} & & & \\                                   
    & +\name{} knowledge & +10.4 & +19.7 & +2.9 & $-$3.3 & $-$10.6 & +4.2 \\                           
    & & \emph{\small 100\% valid} & \emph{\small 100\% valid} &\emph{\small 100\% valid} & & & \\ 
    & \textbf{Gain} & +1.4 & +3.2 & +6.1 & \textbf{+50.9} & \textbf{+45.8} & \textbf{+58.9} \\
  \midrule                                                                                               
  \multirow{2}{*}{\shortstack[l]{\emph{E2ETune}\\\emph{(LLM} \\\emph{one-shot} \\\emph{inference)}}}                                   
    & Original {(PG docs)} & $-$1.0 & +0.7 & +0.0 & +6.9 & +12.0 & +33.6 \\                    
    & +\name{} knowledge & +1.0 & +0.0 & +0.5 & +8.4 & +12.7 & +35.2 \\  
    & \textbf{Gain} & +2.0 & $-$0.7 & +0.5 & \textbf{+1.5} & \textbf{+0.7} & \textbf{+1.6} \\
  \bottomrule                                                                                            
  \end{tabularx}                                                                                       
\end{table*}

Table~\ref{tab:main-results} reports end-to-end results on two testbeds: (1) the \emph{VM cluster}, where profiling conditions match deployment, and (2) the \emph{server}, where profiling knowledge is transferred.

\heading{Matched environment (VM cluster).}
In this setting, all methods run on hardware consistent with the profiling environment. Three patterns emerge from the VM results.

First, traditional documentation rules hurt analytical workloads. PG-Official degrades TPC-H by 54.7\%, confirming the staleness finding from \S\ref{sec:gap:staleness}.
Second, \name eliminates invalid configurations. GPTuner-Original wastes 19--32\% of its 100-trial budget on crashed or severely degraded configurations, while GPTuner+\name{} achieves 100\% valid trial rate across all workloads.
The throughput improvement on TPC-C-w (+19.7\%) is mainly attributable to knowledge quality, where the algorithm and budget are identical.
Third, \name Agent, which executes the full procedural document (sensitivity scan $\to$ correlation detection $\to$ joint optimization), achieves the largest gains by discovering workload-specific configurations through systematic profiling.

\heading{Cross-hardware transfer (server).}
We test whether \name{} knowledge profiled on 8\,GB VMs generalizes to a 192-core, 1\,TB server, a 96$\times$ core-count and 128$\times$ memory gap.

Even across a 96$\times$ hardware gap, \name{} knowledge rescues GPTuner from catastrophic failure. GPTuner-Original, which consumes raw PG documentation via LLM extraction, \emph{halves} throughput on TPC-C-r ($-$54.2\%) and more than doubles latency on TPC-H ($-$54.7\%). Replacing the knowledge source with \name{}'s procedural document, 15 sensitivity-ranked parameters with empirical safe ranges, transforms GPTuner from catastrophic to near-default, with zero changes to its SMAC optimizer.

For E2ETune, which already embeds implicit PostgreSQL knowledge through training, \name{}'s explicit procedural knowledge provides complementary gains (+1.5\% on TPC-C-r, +1.6\% on TPC-H), confirming that the two knowledge forms are not redundant.

\heading{Limitations on TPC-C-w.}
\name{} knowledge does not improve TPC-C-w on server (GPTuner+\name{}: $-$10.6\%). The 1\,TB server has fundamentally different memory pressure for write-heavy workloads than the 8\,GB VMs where profiling was conducted.
This underscores that \name{}'s \emph{reference data} (safe ranges, interaction strengths) is hardware-specific, while the \emph{procedural methodology} (sensitivity scan $\to$ correlation screen $\to$ joint optimize) remains valid. On the VM cluster, where profiling matches deployment, this limitation does not arise.

\subsection{Why Knowledge Form Matters}
\label{sec:eval:form}

We inject five knowledge variants into E2ETune, controlling information quantity while varying the knowledge \emph{form}: whether the LLM receives recommended values (declarative) or structured tuning methodology (procedural).
Table~\ref{tab:knowledge-form} reveals several counter-intuitive patterns:

\begin{table}[htbp]
\centering
\caption{Knowledge form experiment (server, E2ETune, 5 runs per condition). Performance improvement (\%) over default. \emph{Declarative} knowledge provides specific parameter values; \emph{procedural} knowledge provides tuning methodology. Even correct values hurt; procedural structure alone helps.}
\label{tab:knowledge-form}
\small
\begin{tabularx}{\columnwidth}{%
  >{\raggedright\arraybackslash}p{5cm}
  >{\raggedright\arraybackslash}p{2.5cm}
  >{\centering\arraybackslash}X
  >{\centering\arraybackslash}X
  >{\centering\arraybackslash}X
}
\toprule
\textbf{Knowledge} & \textbf{Form} & \textbf{TPC-C-r} & \textbf{TPC-C-w} & \textbf{TPC-H} \\
\midrule
E2ETune Vanilla & None & +6.9 & +12.0 & +33.6 \\
\midrule
\multicolumn{5}{@{}l}{\emph{Declarative (what to set):}} \\
\quad + Wrong Numbers & Incorrect values & +0.2 & +1.7 & $-$8.3 \\
\quad + Full Numbers & Correct values & $-$0.4 & $-$1.9 & $-$2.1 \\
\quad + HW-Relative & Adapted ratios & +1.0 & $-$1.4 & +2.2 \\
\midrule
\multicolumn{5}{@{}l}{\emph{Procedural (how to determine):}} \\
\quad + Structure-Only & Steps, no data & +1.3 & $-$0.1 & +5.3 \\
\quad + Full \name{} & Steps + ref.\ data & \textbf{+8.4} & \textbf{+12.7} & \textbf{+35.2} \\
\bottomrule
\end{tabularx}

\end{table}

\begin{myitemize}
\item \textbf{Declarative knowledge usually fails.}
Even \emph{correct} specific values (Full Numbers) degrade performance on all three workloads ($-$0.4\% to $-$1.9\%) relative to providing no knowledge at all.
The LLM anchors on the injected numbers and reduces its exploration, even when those numbers are suboptimal for the target hardware.
Incorrect values (Wrong Numbers) are catastrophic on TPC-H ($-$8.3\%), directly illustrating the staleness problem identified in \S\ref{sec:gap:staleness}.
\item \textbf{Procedural structure alone outperforms all declarative forms.}
Structure-Only provides only the tuning \emph{steps} (``sweep the parameter; record the response curve; check for correlations'') with no specific data, which already outperforms the best declarative variant on TPC-H (+5.3\% vs.\ +2.2\%).
The \emph{process} carries value independent of specific calibration data.
\item \textbf{Full \name combines structure with data for the best results.}
Full \name{} (procedural steps + empirical reference data) achieves the highest improvement across all workloads.
The reference data accelerates convergence, telling the agent \emph{where} to look, but the procedural structure is the essential ingredient that enables correct reasoning about the parameter space.
\end{myitemize}

The above results validate our key point: documentation should encode \emph{how to determine the right configuration}, rather than \emph{what the right configuration is}.


\subsection{Profiling Cost Analysis}
\label{sec:eval:cost}

\begin{table}[htbp]
\centering
\caption{Cost breakdown. The offline phase runs once per system version; the online phase runs per deployment.}
\label{tab:cost}
\small
\begin{tabularx}{\columnwidth}{%
  >{\raggedright\arraybackslash}p{5cm}
  >{\raggedright\arraybackslash}X
  >{\raggedright\arraybackslash}X
  >{\raggedright\arraybackslash}p{4cm}
}
\toprule
\textbf{Phase} & \textbf{Runs} & \textbf{\%} & \textbf{Parallelism} \\
\midrule
\multicolumn{4}{@{}l}{\emph{Offline (one-time):}} \\
\quad 1. Sensitivity scan & 6{,}297 & 57 & Fully parallel \\
\quad 2. Correlation screen & $\sim$3{,}500 & 32 & Per-pair parallel \\
\quad 3. Joint optimization & $\sim$1{,}200 & 11 & Per-component parallel \\
\quad \textbf{Total} & \textbf{$\sim$11{,}000} & & \\
\midrule
\multicolumn{4}{@{}l}{\emph{Online (per deployment):}} \\
\quad Agent-guided tuning & 30--120 & & Sequential \\
\bottomrule
\end{tabularx}
\end{table}

\begin{figure}[htbp]
  \centering
  \includegraphics[width=0.7\columnwidth]{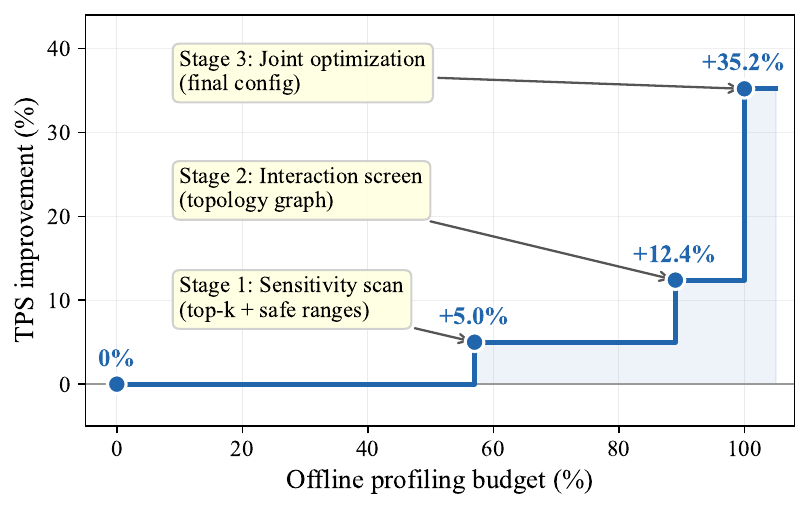}
  \caption{Profiling cost vs.\ tuning quality. Each stage completion triggers a discrete jump in performance.}
  \label{fig:cost-tradeoff}
\end{figure}

Offline profiling ($\sim$11{,}000 runs) decomposes into three stages, each for a \emph{distinct capability} (Table~\ref{tab:cost}). Figure~\ref{fig:cost-tradeoff} shows a clear staircase pattern: performance improves only at stage boundaries, as each stage produces a qualitatively new artifact (parameter ranking, correlation topology, or joint configuration).

\begin{myitemize}
\item \textbf{Discrete returns.}
Profiling exhibits \emph{stage-level gains}: Stage~1 yields +5\% via dimensionality reduction and safe ranges; Stage~2 improves to +12.4\% by exposing correlation structure; Stage~3 reaches +35.2\% through joint optimization.
Partial completion within a stage provides little benefit, as incomplete artifacts (e.g., partial correlation graphs) cannot guide optimization reliably.
\item \textbf{High return under partial investment.}
Even Stage~1 alone (57\% of budget) delivers +5\% by reducing the search space (116 $\rightarrow$ 15 parameters) and eliminating unsafe configurations that waste trials.
\item \textbf{Disproportionate payoff of joint optimization.}
Stage~3 consumes only 11\% of the budget but yields the largest gain (+22.8\%).
This is enabled by topology-aware decomposition from Stage~2, which reduces the search to small interacting groups.
\end{myitemize}

\subsection{Ablation: Contribution of Each Technique}
\label{sec:eval:ablation}

We inject five variants of \name{} knowledge into E2ETune on the high-end server (3 repetitions per condition) to isolate the contribution of each component.

\begin{table}[h]
\centering
\small
\setlength{\tabcolsep}{3pt}
\caption{Knowledge component ablation (high-end server, E2ETune, 3 repetitions, live BenchBase). Performance improvement (\%) over default. Best per workload in \textbf{bold}.}
\label{tab:ablation}
\begin{tabularx}{\columnwidth}{%
  >{\raggedright\arraybackslash}p{7cm}
  >{\centering\arraybackslash}X
  >{\centering\arraybackslash}X
  >{\centering\arraybackslash}X
}
\toprule
\textbf{Condition} & \textbf{TPC-C-r (\%)} & \textbf{TPC-C-w (\%)} & \textbf{TPC-H (\%)} \\
\midrule
E2ETune Vanilla & +6.9 & +12.0 & +33.6 \\
\midrule
+ Ranking only & \textbf{+9.5} & +9.0 & +30.9 \\
+ Interaction only & +3.3 & \textbf{+13.2} & +30.1 \\
+ Safe range only & +7.9 & +12.6 & +32.8 \\
+ Full (\name{}) & +8.4 & +12.7 & \textbf{+35.2} \\
\bottomrule
\end{tabularx}
\end{table}

Table~\ref{tab:ablation} shows that each component's contribution is \textbf{workload-dependent}:
\emph{Ranking dominates on read-heavy OLTP} (TPC-C-r: +9.5\%), because read-heavy workloads are governed by a few memory/I/O parameters that ranking correctly identifies.
\emph{Correlation knowledge dominates on write-heavy OLTP} (TPC-C-w: +13.2\%), because write-heavy workloads involve coupled I/O paths (buffer pool $\leftrightarrow$ WAL $\leftrightarrow$ checkpoint) where one parameter's optimal value depends on another.
\emph{Full knowledge is the most robust}: it achieves the best result on TPC-H (+35.2\%) and near-best on both OLTP workloads, without requiring the user to know which component matters.

To demonstrate the concrete cost of ignoring parameter correlations, we select the \sbuf{} $\times$ \cct{} 
pair and
compare independent and joint optimization on the VM cluster. Tuning each parameter independently to its optimum 
and combining the results yields a \textbf{$-$13.4\%} throughput change on 
TPC-C-r; tuning them jointly over the 2D space yields \textbf{+35.2\%}. This is because a large buffer pool paired with 
infrequent checkpoints causes dirty-page accumulation and I/O storms, but 
each parameter appears optimal in isolation when the other is held at its 
default. This result is the most direct evidence for topology discovery's 
value: the \emph{same parameter values} produce a 48.6\% performance 
gap depending solely on whether the tuning \emph{process} accounted for 
their correlation.

\section{Discussion}

\paragraph{Generalization beyond databases.}
Although \name is evaluated on PostgreSQL, its methodology is not specific to DBMS. Many systems, such as operating systems, language runtimes, Web servers, and distributed systems, also expose large configuration spaces with workload-dependent and cross-parameter effects. In principle, \name can be instantiated for these systems through a similar two-phase workflow: offline profiling discovers sensitivity and interaction structure, while online execution applies the generated skills for deployment-specific tuning. However, different systems have different failure modes and observability constraints; for example, database misconfiguration usually leads to performance degradation, whereas OS-level misconfiguration may cause crashes or service instability. As such, applying \name beyond DBMS may require stronger rollback, sandboxing, and safety-aware recovery mechanisms. \name can be viewed as a general methodology for constructing executable tuning knowledge.

\paragraph{Profiling cost and transferability.}
\name does not eliminate offline profiling overhead. The offline phase requires substantial benchmark execution to construct sensitivity rankings, safe ranges, and correlation graphs; in our PostgreSQL study, this amounts to roughly 11K runs. Since most runs are embarrassingly parallel and performed once per software version, the cost can be amortized across many deployments, making \name more suitable for mature, stable software stacks where tuning is revisited periodically than for one-off or rapidly evolving environments. Meanwhile, the workflow itself, including sensitivity scan, interaction screening, and component-wise joint optimization, is more transferable than the empirical reference data. Safe ranges, sensitivity strengths, and interaction weights may still shift with hardware, workload mix, and deployment scale. Therefore, when the target platform differs substantially from the profiling environment, \name's workflow remains applicable, while part of the reference data should be recalibrated.

\paragraph{Workload dependence.}
\name assumes that offline profiling workloads are representative of the target deployment. This assumption is necessary because tuning decisions are inherently workload-dependent: a parameter that is important for one workload may be irrelevant or even harmful for another. \name partially mitigates this issue by recording per-workload sensitivity profiles and by verifying key decisions during online execution. Nevertheless, it cannot guarantee correctness under arbitrary workload drift. If the online workload activates a previously irrelevant subsystem, or if the workload mix changes substantially over time, the generated skills may focus on an incomplete parameter set. In such cases, workload monitoring and periodic re-profiling are necessary to keep the procedural document aligned with the deployed system.

\paragraph{Higher-order parameter interactions.}
\name currently focuses on pairwise and low-dimensional parameter correlations. This design choice is motivated by practicality: exhaustive exploration of high-order interactions is computationally prohibitive, while our empirical results suggest that many important dependencies can already be captured by sparse, low-dimensional correlation components. However, real systems may contain higher-order effects that are not visible from pairwise screening alone. Consequently, \name may underestimate dependencies that emerge only when three or more parameters change together. A promising direction is hierarchical interaction discovery, where higher-order candidates are explored only when their lower-order subsets exhibit high sensitivity or strong interaction.

\paragraph{Agent reliability.}
The skill-based design reduces ambiguity for LLM-based tuning agents, but it does not fully solve agent reliability. In full procedural execution, the agent must still invoke benchmarks, parse logs, compare measurements, and apply decision rules correctly. Although the generated skills contain preconditions, postconditions, and reference data for self-checking, they cannot completely prevent agent-side failures such as misinterpreting noisy measurements, prematurely terminating experiments, or incorrectly applying a decision rule. Therefore, \name should be understood as a structured control layer that improves the reliability of agentic tuning, not as a complete guarantee of autonomous correctness.

\paragraph{Documentation and skills.}
The goal of \name is not to replace traditional documentation. Static documentation remains valuable for explaining parameter semantics, legal value ranges, and operational cautions. The limitation is that such documentation usually records what experts concluded, but not how they reached those conclusions. \name complements documentation by turning tuning methodology into executable, verifiable skills. This distinction also clarifies the scope of our contribution: \name is not a universal optimizer that guarantees globally optimal configurations. Rather, it provides a knowledge representation that helps agents perform targeted measurement, avoid stale or unsafe recommendations, and account for parameter interactions that static documentation often omits.
\section{Related Work}
\label{sec:related}

\heading{LLM-based system tuning.}
GPTuner~\cite{gptuner} pioneered the use of LLMs to extract structured knowledge
from database documentation and feed it to a Bayesian optimizer. D-Bot~\cite{dbot}
and DB-GPT~\cite{dbgpt} extend this direction by applying LLMs to anomaly diagnosis
and root-cause analysis. Recent systems such as BYOS~\cite{byos}, OS-R1~\cite{os-r1}, and Wayfinder~\cite{wayfinder} further apply LLMs to operating-system configuration tuning and specialization. While these systems demonstrate the potential of LLMs for system optimization, they largely treat the underlying knowledge source as fixed. \name{} is complementary: rather than merely \emph{consuming} existing documentation, it encodes empirically grounded \emph{procedural knowledge} into documentation. The resulting gains transfer across both GPTuner, a BO-based tuner, and E2ETune~\cite{e2etune}, an LLM-inference-based tuner, confirming that the improvement lies in the upgraded knowledge rather than in a particular tuning algorithm.

\heading{ML-based database tuning.}
OtterTune~\cite{ottertune} uses Gaussian processes with workload mapping and LASSO-based parameter selection, implicitly assuming that importance decomposes into independent per-parameter contributions. CDBTune~\cite{cdbtune} applies deep RL and can in principle capture correlations, but requires thousands of trials to do so. ResTune~\cite{restune} ensembles multiple GP models and LlamaTune~\cite{llamatune} reduces dimensionality via random projection; neither explicitly models interaction structure. Hunter~\cite{hunter} transfers tuning knowledge across workloads, assuming that parameter importance is portable across contexts. These systems focus on improving the \emph{search algorithm} while treating the knowledge source as fixed. \name{} is orthogonal: it improves the \emph{knowledge} that feeds any such algorithm, as our controlled experiments (\S\ref{sec:eval}) confirm.

\heading{Configuration correlation analysis.}
ConfigCrusher~\cite{configcrusher}, ConEx~\cite{conex}, and SPLConqueror~\cite{splconqueror} characterize parameter correlations for debugging and understanding but do not translate findings into tuning guidance. \name{} closes this gap: it uses factorial ANOVA for correlation screening and encodes the discovered topology into executable skills that drive an optimization agent.


\heading{Operational knowledge management.}
The SRE community has developed runbooks, playbooks, and incident response procedures that encode operational knowledge in semi-structured form~\cite{sre-book}.
Recent work explores LLM-based runbook execution~\cite{runbook-llm}.
These efforts share \name{}'s philosophy of procedural knowledge but target incident response rather than performance tuning, and lack the calibration that grounds \name{}'s decision criteria. 


\section{Conclusion}
\label{sec:conclusion}

Today's system tuning guides capture what experts conclude but discard how 
they reason. We argue for a paradigm shift: documentation should encode 
\emph{how to determine the right configuration}, not merely \emph{what the right configuration is}.
\name is a tool that translates tuning methodologies into actionable skills, empowering LLM-based agents for efficient and effective tuning.
We believe grounding tuning knowledge in reproducible process is the essential first step toward agentic tuning.

\bibliographystyle{plainnat}
\bibliography{references}



\end{document}